\documentclass[graybox]{svmult}

\usepackage{type1cm}        
\usepackage{makeidx}         
\usepackage{graphicx}        
\usepackage{multicol}        
\usepackage[bottom]{footmisc}

\usepackage{newtxtext}       %
\usepackage{newtxmath}       

\usepackage[utf8]{inputenc}

\makeindex             
\def\mnras{Monthly Notices of the Royal Astronomical Society, }
\def\apj{Astrophysical Journal, }
\def\aplett{Astrophysical Letters, }
\def\apjl{Astrophysical Journal Letters, }
\def\aap{Astronomy and Astrophysics, }
\def\aar{Astronomy and Astrophysics Review, }
\def\physrep{Physics Reports, }

\def\nat{Nature, }

\def\aj{Astronomical Journal, }

\begin{document}

\title{Radio Millisecond pulsars}

\author{Bhaswati Bhattacharyya and Jayanta Roy}
\institute{Bhaswati Bhattacharyya \at National Centre for Radio Astrophysics, Tata Institute of Fundamental Research, Pune 411 007, India, \email{bhaswati@ncra.tifr.res.in}
\and Jayanta Roy \at National Centre for Radio Astrophysics, Tata Institute of Fundamental Research, Pune 411 007, India, \email{jroy@ncra.tifr.res.in}}
%
%
\maketitle
\label{ch:bhaswati}
\abstract{The extreme timing stability of  radio millisecond pulsars (MSPs) combined with their exotic environment and evolutionary history makes them excellent laboratories  to probe matter in extreme condition. Population studies indicate that we have discovered less than five per cent of the MSPs of our Galaxy, implying that a huge majority of radio MSPs are waiting to be discovered with  improved search techniques and more sensitive surveys. In this chapter, we provide an overview of the present status of ongoing and upcoming surveys for MSPs. Observed spectra, profile and polarisation properties of known radio MSPs are also summarised. 
Finally, we describe how the timing studies of radio MSPs enable a huge science return including attempts to detect gravitational waves using an array of MSPs, gravity tests using individual interesting MSP systems, as well as probing the  intra-binary material using eclipses observed in MSPs in compact binary systems.}

\section{Introduction to parameters of radio MSPs}
\label{sec:bh1}
Millisecond pulsars (MSPs) are rapidly rotating neutron stars (rotational period of few tens of milliseconds) with very small spin-down rates.
\begin{figure}[h]
  \includegraphics[width=6.5in,angle=0]{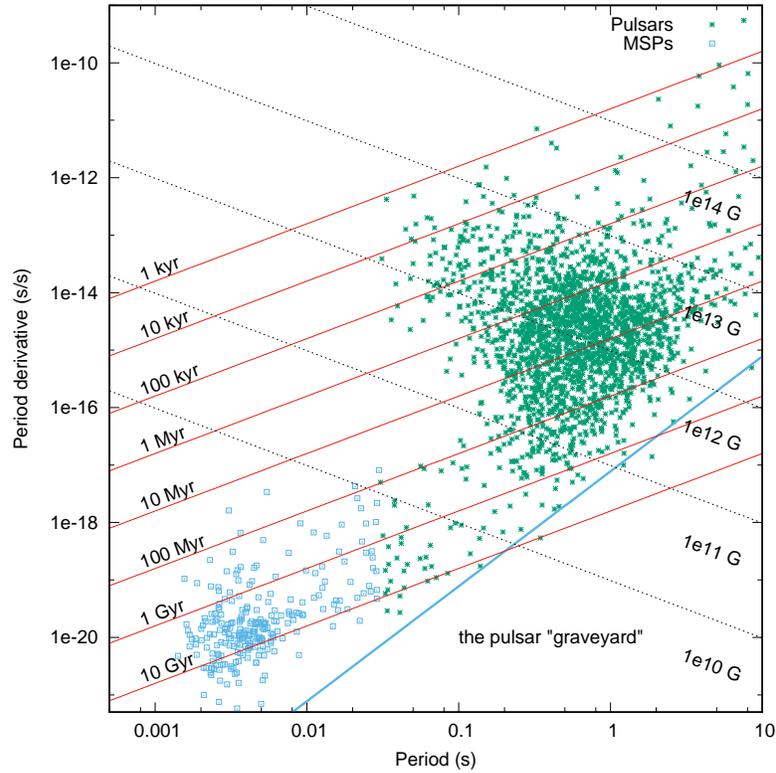}
\caption{Period versus period derivative of the {\it ordinary} ($P>30$~ms)  pulsar and the millisecond pulsars.}
\label{fig_p_pd}
\end{figure}
 Whereas the spin period of the radio pulsars span around four orders of magnitude (1.4 ms to 23 s), MSPs are defined here by a periodicity $<$ 30 ms. With extremely stable periods and very low period derivative values, MSPs are the most precise celestial clocks and occupy the bottom-left corner in the $P-\dot{P}$ diagram (see Fig.~\ref{fig_p_pd}, where  blue squares mark MSPs; see also see also Fig. 4.1 of Chapter~4). Since rotation-powered pulsars spin down at a rate which depends on the magnetic field strength and the spin period of the pulsar, it is inferred that the magnetic field of an MSP is a few orders of magnitude weaker than an {\it ordinary}, slower pulsar. MSPs are assumed to have acquired their high rotational rate by accretion of matter, and thereby transfer of angular momentum, from a low mass ($<M_{\odot}$) companion star in a binary system \cite{alpar82,bh91}.
  Mass accretion is also possibly responsible for the decay of the magnetic field of the neutron star.
 Such a recycling scenario (discussed in Sect.~4.1.2 of Chapter~4)  is now supported by observational evidences of accreting millisecond pulsars (see Chapter~4) and of a few transitional systems switching states between radio MSP and low-mass X-ray binaries (see Chapter~6).
\begin{table*}
\begin{center}
\caption{Parameters of the known radio MSPs}
\vspace{0.3cm}
\label{msp_parameters}
\begin{tabular}{|l|c|c|c|c|c|c|c|c|c|c|c|c|c|c|c|c}
\hline
Parameters  & Range of values (units) \\
&\\\hline
Spin Period ($P$) & 1.4$-$30 ms \\
                  &\\
Spin Period derivative ($\dot{P}$)   & 10$^{-18}-$ 10$^{-22}$ \\
                  &\\
Magnetic field strength (B) & 10$^7-$10$^9$ G\\      
                  &\\
Age & 10$^7-$10$^{11}$ years\\
                  &\\
Dispersion Measure (DM) & 2.6$-$540 pc~cm$^{-3}$    \\
                  &\\
DM distance (d)&  0.11$-$49 kpc \\
                  &\\
Flux density (S$_{1400}$) & 0.01 $-$ 150 mJy\\
                  &\\
Companion mass (M${_c}_{min}$) & 0.009$-$1.39 M$_{sun}$  \\
                  &\\
Orbital period (P$_b$)& 0.065$-$669 days  \\
                 &\\
Eccentricity (e) & 0$-$0.95    \\
                 &\\
Semi-major axis (A$_1$) &  0.0018$-$100 lt-sec\\
&     \\\hline
\end{tabular}
\end{center}
\vspace{1cm}
\end{table*}

 MSPs are still a small population  compared to the classical {\it ordinary} pulsars (spin period $>$ 30 ms). A total of 512 MSPs are reported in the lists maintained by E.~Ferrara and D.~Lorimer\footnote{Available at \url{http://astro.phys.wvu.edu/GalacticMSPs/}}. and by P.~Freire\footnote{{Available at \url{http://www.naic.edu/~pfreire/GCpsr.html}.}}, as of November 2020, whereas 2450 slower pulsars are listed in the Australia Telescope National Facility (ATNF) database \cite{manchester2005}\footnote{Available at \url{https://www.atnf.csiro.au/research/pulsar/psrcat/}.}.
 The parameters of the  MSPs listed in the ATNF pulsar catalogue are summarised in Table \ref{msp_parameters}. The recycling scenario of MSPs suggests that most of them should be part of a  binary system, and this is realized in $>$80\% of the known systems  
 having a companion star. 
 MSPs in binary systems are found in systems with a period ranging from 75 minutes to 669 days, and a mass of the companion star between 0.009 and 1.39 M$_{\odot}$ (see also Fig.~6.1 of Chapter~6). More than  60\% are found in systems with an orbital period P$_b$ $>$ 1 day. These large-period binary MSPs essentially fall into three groups, depending on
the mass of the companion star. The majority ($\sim$85\%) have a low-mass ($<$ 0.4 M$_{\odot}$)  Helium white dwarf companion, but 
higher-mass (seemingly Carbon$-$Oxygen) white dwarfs, as well as neutron star companions are also found.  The mass of the companion increases as a function of orbital period, following theoretical expectations \cite{podsiadlowski02}. Most of the MSPs are in nearly circular orbit with only $\sim$8\% of known binaries having known eccentricity value $>$ 0.1.  Tauris \& Savonije \cite{Tauris99} and Hui et al. \cite{Hui18} analysed the observed orbital properties of a binary MSPs with a white dwarf companion and reported a positive correlation between the orbital period and the  eccentricity. However, different trends were obtained for MSPs with a Helium white dwarf companion, and MSPs with a Carbon$-$Oxygen white dwarf. They also reported two gaps in the 
distribution of orbital period (between 35$-$50 days and between 2.5$-$4.5 days). On the other hand, 
MSPs in binaries with a short orbital period (P$_b$ $<$ 1 day) also include eclipsing pulsars (see Sect.~\ref{subsec:bh6.3}) either with a non degenerate main-sequence companion with a mass in the range $0.1$--$0.8$  M$_{\odot}$ (dubbed {\it redbacks}) or with a < 0.06 M$_{\odot}$ brown dwarf (termed {\it black widows}). The reader is referred to the Chapter~7 for a detailed discussion of the origin and evolutionary channels of MSPs.

Estimates for the Galactic population of MSPs range from 40,000 to 90,000 objects \cite{faucher06}, indicating that a large number of MSPs are waiting to be discovered. Presently, only $\sim$15\% of the $\sim$ 3000 known pulsars are MSPs, either in the Galactic disk or in globular clusters. Thus, it is possible that the known parameter range of the MSPs does not represent the true distribution. 

This chapter presents a overview of radio millisecond pulsars. 
Sect.~\ref{sec:bh2} of this chapter details spectra and polarisation properties of the MSPs.
Searches for radio MSPs are detailed in Sect.~\ref{sec:bh3}. Some aspects of the timing studies of radio MSPs are presented in Sect.~\ref{sec:bh4}. 

\section{Properties of MSPs}
\label{sec:bh2}
\subsection{Spectra and luminosity}
\label{subsec:bh2}
\label{subsec:bh2.1}
\begin{figure}[t!]
\includegraphics[scale=.4]{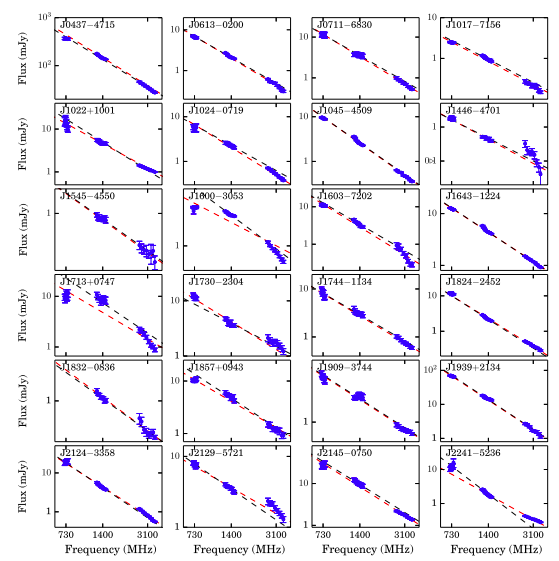}
\centering
\caption{Flux density spectra for 24 MSPs (Credit \cite{dai15}), the red and black lines are for spectral index spectra fitted with power law $\alpha_1$ and $\alpha_2$.}
\label{fig_spec_MSP}
\end{figure}

Kramer et al. \cite{kramer98} compared the spectra of  {\it ordinary} (P>30~ms) pulsars and MSPs observed in the 0.7--3.1 GHz band.
They concluded that the average spectra of MSPs are steeper than  {\it ordinary} pulsars. They derived a mean spectral index of $-$1.8 $\pm$ 0.1 for a set of 32 MSPs located in the Galactic disk and a mean index of $-$1.60 $\pm$ 0.04 for {\it ordinary}  pulsars in the same frequency range. The median values for both samples are $-$1.8 and $-$1.7, respectively. However, they  also pointed out that the steeper spectral index for MSPs  could be due to the selection bias of having fainter (and farther) {\it ordinary}  pulsars in the sample, with a relatively flatter spectral index. Indeed, restricting the  data set to sources that are closer than 1.5 kpc, they found that mean spectral index of MSPs and {\it ordinary}  pulsars are similar ($-$1.6 $\pm$ 0.2 for MSPs and $-$1.7 $\pm$ 0.1 for {\it ordinary}  pulsars), with a median value of $-$1.65 and $-$1.66, respectively. 
Note that a more recent study by Bates et al. \cite{bates14} based on  larger sample of {\it ordinary}  pulsars reported that the distribution of the spectral index has a mean of $-1.4$ and a standard deviations of $1.0$. Although the number of MSP has increased drastically in last two decades
since the study by Kramer et al.\cite{kramer98},  the flux at more than one observing frequency was reported only for only a small
fraction of the newly discovered MSPs. In a more recent study, Dai et al. \cite{dai15} investigated 24 MSPs observed with the Parkes 64-m telescope in three bands, centred at 730, 1400 and 3100 MHz.
Fig. \ref{fig_spec_MSP} plots the flux density spectra of these  MSPs. They reported that the spectra of a few pulsars  significantly deviated from a single power-law across the observing bands. Although a spectral steepening at high frequencies was observed for a few MSPs, for some other a spectral flattening  was instead observed. Dai et al. \cite{dai15} also studied the pulse phase-resolved spectral index of MSPs and found that different profile components have different spectral indices which overlap with one another.
We conclude that considering the observed diversity of the spectral properties of MSPs, and in the absence of systematic flux measurements for a large sample of MSPs, it is not possible to draw a firm conclusion from the comparison of the steepness of the spectra of MSPs and {\it ordinary}  pulsars.

{\it Ordinary}  pulsars exhibit low-frequency turn overs in the spectra, while this is still debated for MSPs. Kuzmin et al. \cite{kuzmin01} analysed the spectra of 30 MSPs  down to 100 MHz, and most of them did not exhibit any low frequency turn over.
On the other hand, Kunyoshi et al. \cite{kuniyoshi15} found that  one forth of the MSPs for which the spectrum was observed down to  100 MHz showed evidence of a turn over (i.e., 10 out of 39 MSPs).

MSPs tend to be less luminous and less efficient radio emitters compared
to {\it ordinary}  pulsars. To probe the MSP birth rate it is important to compare the luminosity of MSPs and {\it ordinary}  pulsars.
Kramer et al. \cite{kramer98} calculated the values of 
 $S \times d^2$ ($S$ being the flux density of the MSPs and $d$ being the distance) and found that the luminosity of MSPs are an order of magnitude fainter than {\it ordinary}  pulsars. They also compared the luminosity distribution of {\it ordinary}  pulsars within a distance of 1.5 kpc, and concluded that some high luminosity MSPs (which should be easy to detect) are missing. 
Finally, they also noted that isolated MSPs are generally fainter than the ones in binaries, which could be attributed to different evolutionary history.

\subsection{Pulse profile and polarisation properties}
\label{subsec:bh2.2}
Xilouris et al. \cite{xilouris98} reported that the pulse profiles of MSPs are slightly more complex than {\it ordinary}  pulsars. 
They considered the number of Gaussian components required to represent the pulse profile as a measure of their complexity. They found that MSP profiles could  be fitted with four Gaussian components on average, whereas three components were enough for {\it ordinary}  pulsar. 

{\it Ordinary} pulsars follow a systematic behaviour, where the observed pulse profile becomes narrower at higher frequencies, which is known as the `radius to frequency mapping'  \cite{rankin93,lyne88}.  Xilouris et al. \cite{xilouris98} reported a much less marked dependence on  frequency for MSP profiles, instead. They identified three categories of MSPs: (i) almost no dependence , (ii) a very slow `radius to frequency mapping'; and (iii) contrary to `radius to frequency mapping'.
They suggested that the observed profile complexity, including the low-level emission and the unusual features
identified in some of the MSPs, could result from emission in outer gaps \cite{romani95}.
 The sample of MSPs studied by Dai et al. \cite{dai15} also confirmed that most MSPs have very wide profiles with multiple components. The majority of the MSPs in their sample showed a duty cyle higher than 50\%, with the profile components which did not  show an appreciable dependence on the observing frequency.
 
 Whereas the investigation by Dai et al. \cite{dai15} covered the band from 730 MHz up till 3100 MHz in three bands, a more recent study by Kondratiev et al. \cite{kondratiev16} presented a census of MSPs using the LOw-Frequency ARray ({\it LOFAR}) in the frequency range 110$-$188 MHz. They found that the separation between the different components of the profiles seen at low-frequency by {\it LOFAR} was compatible with that seen at higher frequencies. Also the width of the profiles was similar at different frequencies.
  Thus low-frequency observations also supported that there was very little pulse profile dependence on frequency. This is different from the classical pulsars and indicates a more compact emission region in the MSP magnetosphere and possibly higher multipolar components. In addition, the observed pulse shapes indicated that the emission beam of MSPs are narrower than the classical pulsars.
Manchester et al. \cite{manchester95} and Ravi et al. \cite{ravi10} suggested that the features in the radio profiles of the MSPs represent caustics in the emission beam. They proposed that the radio emission of MSPs could be originating in wide beams higher up in the pulsar magnetosphere (up to or even beyond the null charge surface). More details on the physics of the emission of MSPs are discussed in Chapter~3.

Xilouris et al. \cite{xilouris98} also studied for the first time the polarization profiles of MSPs  and found that the the polarization degree is higher than in  {\it ordinary}  pulsars. In addition, the swings of the  polarization position angle of MSPs are flatter than in {\it ordinary}  pulsars. The polarization position angle curves of the MSPs  exhibit smaller excursions and cannot be described by rotating vector model (RVM, \cite{radhakrishnan69}). This warrants different models to  explain the MSP polarization properties. To address this, some of the existing models for MSP polarization
suggested  emission from locations which extend over a substantial fraction of the light cylinder \cite{barnard86}.
In addition, it is possible that special geometries of MSPs in binaries \cite{chen93}, or the existence of higher multipole moments in the  magnetosphere of MSPs \cite{manchester95}, can explain the observed polarization properties for individual MSPs. 

\begin{figure}[th!]
\centering
\includegraphics[scale=.5]{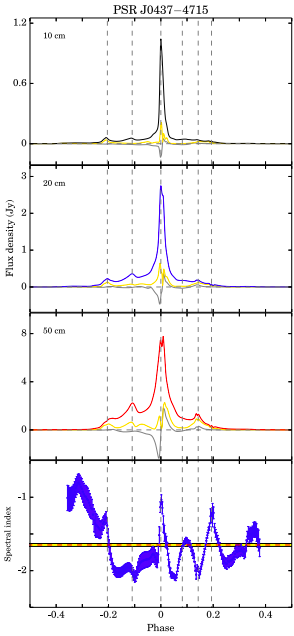}
\caption{The polarization profile of PSR J0437$-$4715 and phase-resolved results. The spectral index observed at different  pulse phases is reported in the bottom panel. The leading and trailing parts have steeper spectral indices, whereas the outer edges of the profile have flatter spectra (Figure Courtesy : \cite{dai15}.)}
\label{fig_pol}
\end{figure}

Dai et al. \cite{dai15} reported that the secondary pre- and post-cursors peaks in the profile  generally have a higher fractional linear polarization than the main pulse. They also observed that the circular polarization showed complicated variations with both frequency and pulse phase, and different pulse components often had different signs of circular polarization. They studied the distributions of the fractional linear  and circular polarization across the frequency bands, finding that although the fractional linear polarization was similar across three bands, both the fractional and net circular polarization decreased at lower frequencies.
 They further reported that the polarization angle sweep for all the MSPs of their sample were extremely complicated and could not be fitted using the RVM. As an example, Fig. \ref{fig_pol} shows the polarization profile or MSP J0437--47. They also noted that  the polarisation angle profile could significantly evolve across the observing frequency band.
 
\section{Searches for Radio MSPs}
\label{sec:bh3}
\begin{figure}[th!]
\centering
\includegraphics[scale=.5]{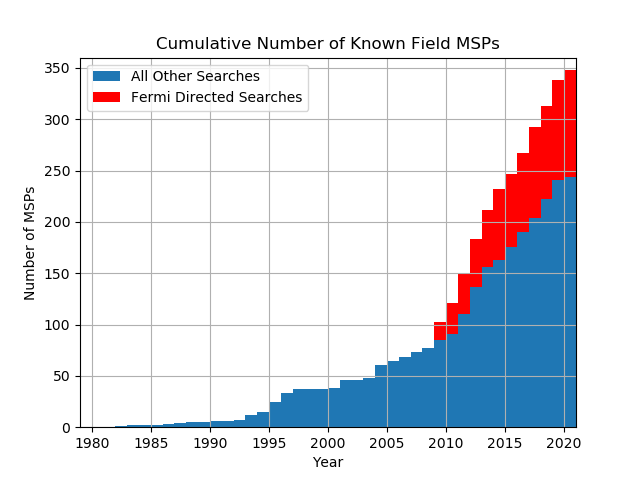}
\caption{Cumulative number of known MSPs in the Galactic field (Figure Courtesy : Paul Ray). Fermi-directed searches have contributed to one-third of this number.}
\label{fig_cum_MSP}
\end{figure}

\begin{figure}[th]
\includegraphics[scale=.7]{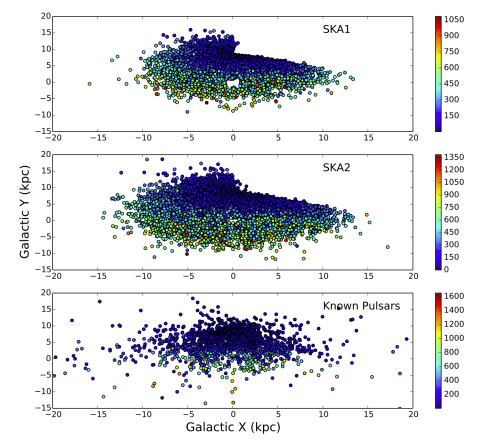}
\caption{{\it PsrPopPy} simulation of the  number of pulsars expected to be found with the {\it SKA} along with their distribution throughout
the Galaxy, projected onto the Galactic plane, compared to the distribution of currently known pulsars. The color coding indicates the 
approximate range of dispersion measures of the simulated pulsars (Figure Courtesy : \cite{Keane15}) }. 
\label{fig_psrpoppy_ska}
\end{figure}

The improvement in the technique of analysis made the rate of discovery of pulsars in ongoing surveys at major telescopes to increase dramatically over the last decade (see Fig. \ref{fig_cum_MSP}). However, the  population of currently known MSPs ($\sim 500$) is less than one per cent of the predicted number of potentially observable 
radio pulsars in the Galaxy ($1.2\times10^5$; \cite{faucher06}). The  {\it PsrPopPy}\footnote{https://github.com/samb8s/PsrPopPy} code is widely used to infer  predictions on the underlying unseen population of  MSPs \cite{bates14}. Once the survey specifications are given as input, the {\it PsrPopPy} simulation can predict the number of MSPs that can be potentially discovered. For example, {\it PsrPopPy} simulations predicted that $\sim$ 3000 MSPs will be discovered by the Square Kilimetre Array ({\it SKA}, \cite{Keane15}; see 
Fig. \ref{fig_psrpoppy_ska}). Thus, population studies indicate that a large population of MSPs are waiting to be discovered. A large fraction of the MSPs are faint sources requiring sensitive searches and improved analysis techniques to be discovered.

The sensitivity of the pulsar surveys is calculated using the radiometer equation. A pulsar will be detectable (with a 5$\sigma$ detection significance) in a survey made of an incoherent array of smaller telescopes, if it exceeds some minimum flux density ($S_{\rm pulsar}$)
that can be calculated using the radiometer equation:
\begin{equation}
S_{\rm pulsar} \sim 5 {{T_{\rm rec}+T_{\rm sky}}\over{G\sqrt{B N_{\rm p} N_{\rm a}t}}}
{\sqrt{{w}\over{P-w}}}
\end{equation}
where $T_{\rm rec}$ and $T_{\rm sky}$ are the temperatures of the receiver and sky respectively, $G$ the gain of individual antennas,
$B$ the bandwidth, $N_{\rm p}$ the number of orthogonal polarizations needed to construct total intensity (i.e. stokes I), $N_{\rm a}$ the number of antennas, $t$ the integration
time, $w$ the effective pulse width (including all instrumental smearing), and $P$ the pulse period.
The limiting sensitivity of different surveys can be calculated using the survey parameters in this equation.
The discovery of new MSPs is hampered  by their radio faintness and requires deeper searches with larger
telescopes. Ongoing searches with the Green Bank Telescope ({\it GBT}), the Parkes telescope, Effelsberg, Arecibo, the Giant Metrewave Radio Telescope ({\it GMRT}), and the Five hundred meter Aperture Spherical Telescope ({\it FAST}) have discovered a good number of MSPs, bringing the total number of MSP in the Galactic field to $\sim$ 353\footnote{\url{http://astro.phys.wvu.edu/GalacticMSPs/GalacticMSPs.txt}}, and total number of MSPs in Globular cluster to 147.
Some of the major radio telescopes that are actively discovering MSPs are large single dish telescopes (e.g. Arecibo, {\it GBT}, Parkes), and  their limiting sensitivity has almost been reached. Thus, large arrays of many smaller telescopes are the future to increase the sensitivity, and this will  ultimately lead to the world's largest telescope, the Square Kilometer Array ({\it SKA}).
\begin{table*}
\begin{center}
\caption{Summary table for ongoing targeted and wide field searches for MSPs}
\vspace{0.3cm}
\label{discovery}
\begin{tabular}{|l|c|c|c|c|c|c|c|c|c|c|c|c|c|c|c|c}
\hline
Telescope  & Survey Name & S$_{min}^\dagger$ & Frequency & MSP discovered & Status \\
           &             &  (mJy)            &  (MHz)    &               &\\\hline

GBT       & Fermi-directed  & 0.06$-$0.08            & 350, 820, 2000$^{\dagger\dagger}$       & 45            & ongoing \cite{Ransom11} \\
Arecibo   & Fermi-directed  & $-$  & 300$-$500, 1214$-$1537 & 14 & ongoing \\
Parkes    & Fermi-directed  & 0.2            & 1262$-$1518  & 18            &  dormant\cite{Camilo15} \\
GMRT      & Fermi-directed  & 0.3$-$0.9            & 306$-$338,607$-$639   & 8           & restarting\cite{bh13}\\
LOFAR     & Fermi-directed  & 1.1          &  115$-$154          & 3             &  ongoing\cite{bassa17,pleunis17} \\
Nancay    & Fermi-directed  & $-$            & 1344$-$1472          & 3            & dormant\cite{Cognard11}\\
Effelsberg & Fermi-directed  & 0.02$-$0.06           &  1180$-$1420         & 1            &  dormant\cite{Barr13}\\ 
FAST      & Fermi-directed  & $-$            &      $-$     & 2            &  ongoing \\
MeerKAT$^{\dagger\dagger\dagger}$ & Fermi-directed & 0.02$-$0.06 & 900$-$1680 & $-$ & starting \\
CHIME$^*$ & CHIME/Pulsar & 0.2 & 400$-$800  &$-$ & starting\cite{Chime20}\\

\hline
Arecibo   & 327 MHz Drift Survey & 0.5 & 300$-$350 & 10 & ongoing$^{a}$\cite{Martinez19} \\
Arecibo  & PALFA & $-$ & 1214$-$1537 & 8 & ongoing\cite{Parent19} \\
GBT       & GBNCC Survey & 0.74 & 300$-$400 & 24 & ongoing$^{b}$\cite{McEwen20} \\
LOFAR     & LOTAAS               & 1.2      & 119$-$151  & 2 &ongoing$^{c}$\cite{Sanidas2019} \\
Parkes    & SUPERB               & 0.2$-$0.7 & 1182$-$1582  & 2 & ongoing$^{d}$\cite{Keane18}\\
GMRT      & GHRSS                & 0.2$-$0.5      & 300$-$500  & 2 & ongoing$^{e}$\cite{bh16,bh19}\\ 
MeerKAT$^{\dagger\dagger\dagger}$ & TRAPUM$-$UHF &  $-$ &544$-$1088  &$-$ & starting\\
\hline
\end{tabular}
\end{center}
$^\dagger$: S$_{min}$ is the limiting flux density at the frequency at which survey is conducted. S$_{min}$ is reported if value is available in related publication/webpage.\\
$\dagger\dagger$: 100, 200, and 800 MHz bandwidth centered at 350, 820, and 1500 MHz\\
$\dagger\dagger\dagger$ : part of TRAPUM project. From private communication with Ben Stappers.\\
$*$: For CHIME/Pulsar the sensitivity limit is for single transit and multi-day stacking search is planned for the targeted survey\\
$a$:https://www.naic.edu/$\sim$deneva/drift-search/ \\
$b$:http://astro.phys.wvu.edu/GBNCC/ \\
$c$:http://www.astron.nl/lotaas/ \\
$d$:https://sites.google.com/site/publicsuperb\\
$e$:http://www.ncra.tifr.res.in/ncra/research/research-at-ncra-tifr/research-areas/pulsarSurveys/GHRSS\\
\vspace{1cm}
\end{table*}

\begin{figure}[t!]
\centering
\includegraphics[width=\textwidth,angle=0]{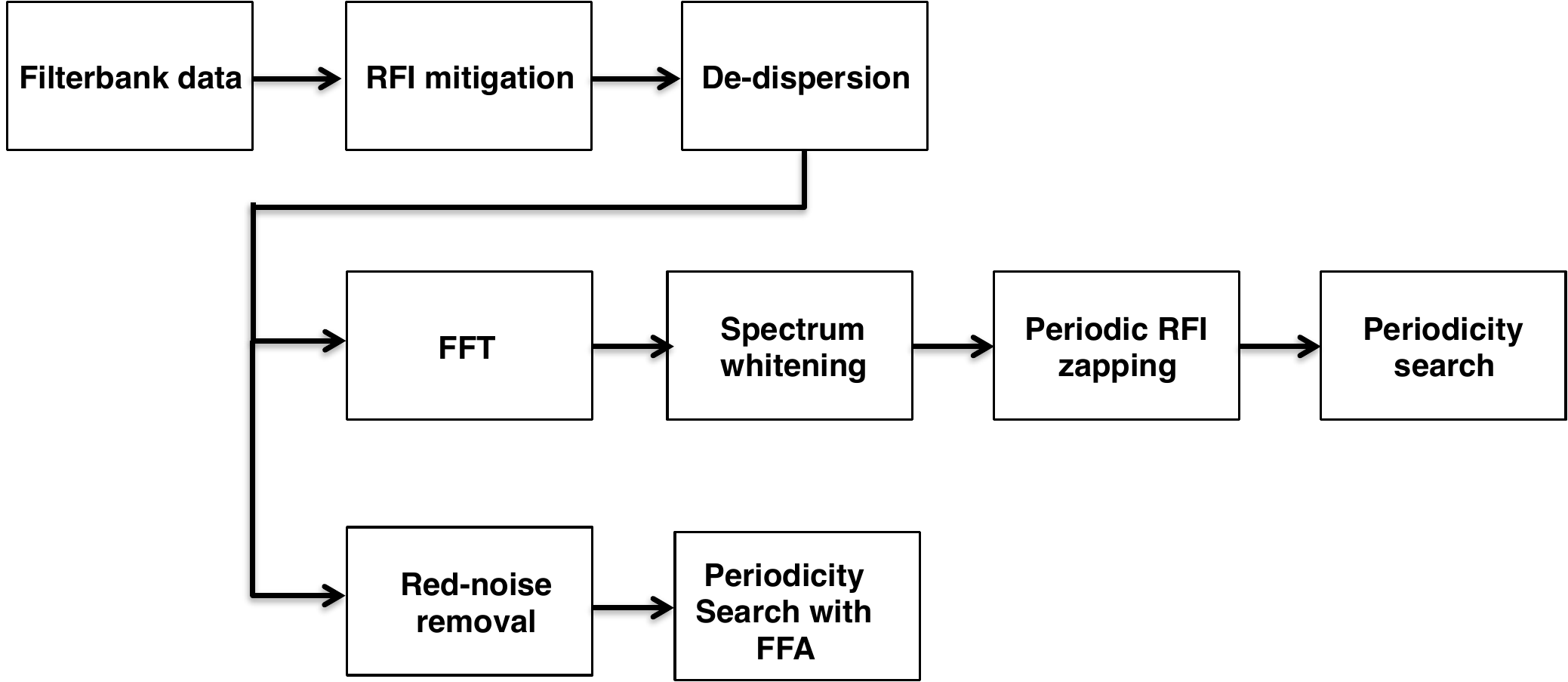}
\caption{Functional blocks for pulsar search processing. The de-dispersed time samples are processed concurrently using frequency-domain search (with Fast Fourier Transforms) and time-domain periodicity search (with Fast Folding Algorithm, FFA).}
\label{fig_block search} 
\end{figure}
In spite of the fact that the rate of discovery of pulsars in ongoing surveys at major telescopes   has increased dramatically over the last decade, the presently known population is a very small fraction of the predicted number of MSPs. Since MSPs are intrinsically faint and most of the MSPs are part of binary systems, a binary acceleration search  (and sometimes jerk search i.e. searching up to period double derivatives) is also required, in addition to a  search for dispersion measure and periodicity. The details of search techniques are described in Sect.~ \ref {subsec:bh3.1}.
 Targeted searches and wide-area blind surveys are two popular ways to look for the large number of the MSPs which are yet unseen. In Sections \ref{subsec:bh3.2}, and \ref{subsec:bh3.3} we describe  these two  MSP search techniques.
 
\subsection{Search techniques}
\label{subsec:bh3.1}
Pulsar search processing is a computing-intensive task. Fig. \ref{fig_block search} shows a typical functional block diagram for a search analysis. The time-frequency filterbank data from the telescope are first processed to excise broad-band and narrow-band radio frequency interference (RFI). RFI mitigated filterbank data are then fed into a de-dispersion transform module, which  corrects for the frequency dependent dispersive delays at various trial dispersion measure (DM) values. 
The pipeline performs a periodicity search for each of the de-dispersed time-series. The periodicity search in frequency-domain involves Fast Fourier Transforms (FFT), spectrum whitening to remove the instrumental red-noise, and masking periodic RFIs (e.g. impulsive signals from AC power-line). In parallel, the de-dispersed time-series can also be searched for periodic signals in the time-domain. 
\begin{figure}[t!]
\centering
\includegraphics[scale=.5,angle=0]{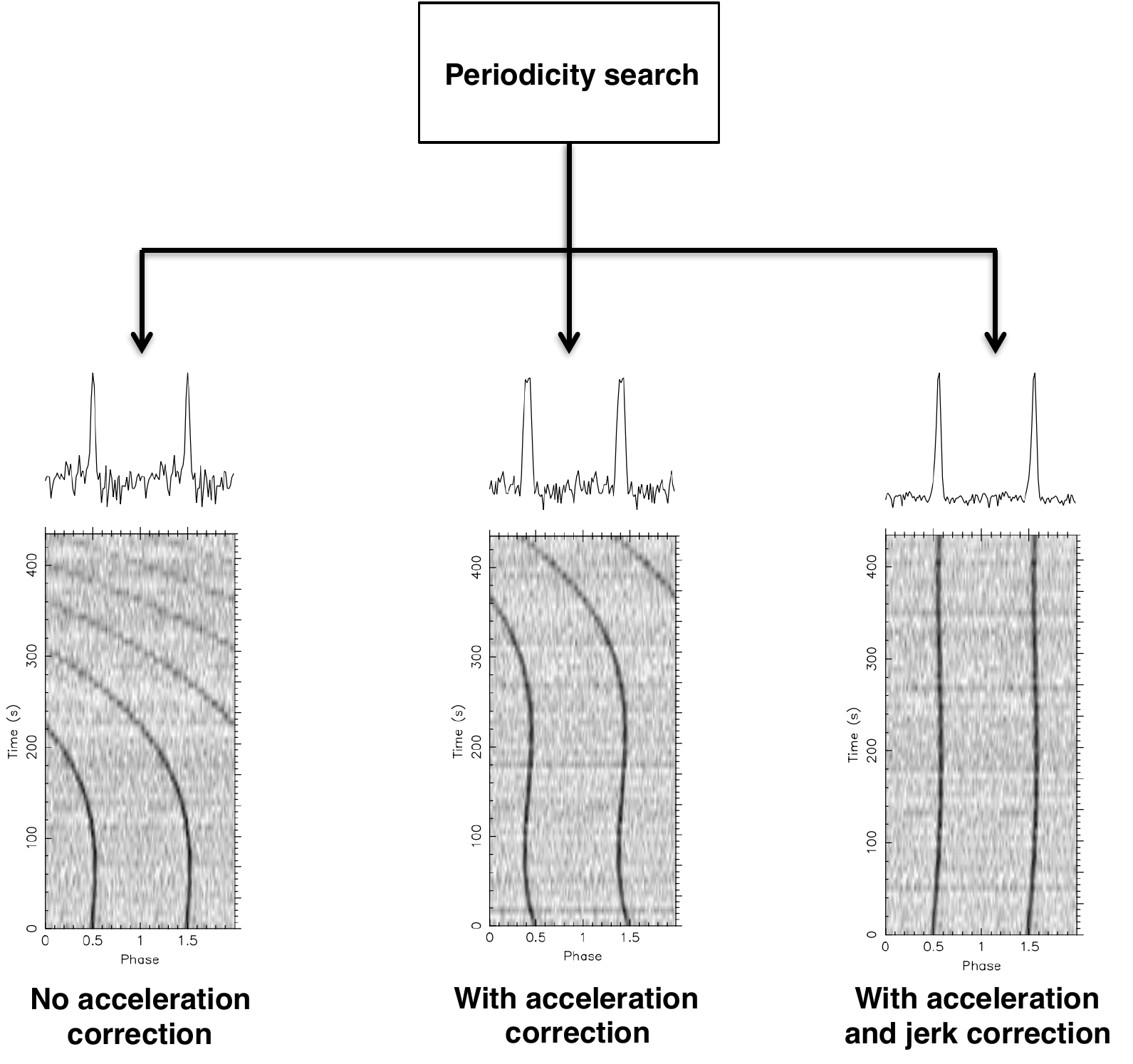}
\caption{The effect of line-of-sight acceleration for a simulated system containing a pulsar in 1.2 hours circular orbit with companion mass of 1.4 M$_{\odot}$. The tracks indicate the pulse intensities as function of time and the averaged pulse profiles are shown on the top.}
\label{fig_periodicity search} 
\end{figure}
The increase in computing power enhanced the sensitivity of on-going surveys with large single dishes or interferometric arrays, making them progress through a hitherto unexplored parameter space. Since the majority of MSPs are in binaries, a periodicity search requires the correction of the line-of-sight acceleration caused by the orbital motion of the pulsar. Thus, in addition to the constant acceleration search, for systems like double neutron star binaries with higher companion mass, the assumption of constant spin frequency-derivative over the span of the observation is no longer valid. The periodicity search employs a jerk search, i.e. searching over the period and the first two period-derivatives that corrects the binary acceleration effect on much shorter time scales even for a fraction of the orbit. Fig. \ref{fig_periodicity search} shows  the improvement obtained in a  periodicity search that involves  no acceleration, a constant acceleration and acceleration+jerk, respectively. The comparison is done for a simulated system containing a 22 ms pulsar and a neutron star companion (1.4 M$_{\odot}$) in a 1.2 hours circular orbit. Correcting for the binary acceleration significantly enhances the signal-to-noise (S/N) of the signal, making a detection much easier. However such an acceleration (and jerk) search increases the pulsar search processing cost by more than an order of magnitude. For example, Anderson et al. \cite{Anderson18} reported an increase  by a factor of $\sim$80 of the processing time while searching for highly accelerated pulsars in Terzan 5 globular cluster. A new 2.93 ms pulsar J1748$-$2446am was discovered with the jerk search which had  not been detected before with an acceleration-only search \cite{Anderson18}. 
Discovering such systems is important as they provide unique laboratories to test the theories of gravity (see Sect. \ref{subsec:bh6.2}).  
The population of non-recycled slow pulsars (period $>$ 100 ms) contains several interesting objects, such as pulsars showing pulse intermittence, drifting and nulling, all of which are important probes of the emission physics. We also know two  ultra-slow pulsars  with a period longer than 10~s; these pulsars graze the theoretical death-line and are interesting to probe the conditions at which the radio emission is expected to cease. 
The instrumental red-noise and radio frequency interference (RFI) reduce the search sensitivity at the low frequency end of the power spectrum of the detected time series, where the signal from these objects is strongest. In addition, due to the shorter duty cycle of long period pulsars, the number of harmonics used in the frequency-domain periodicity search limits the signal recovery from the power spectrum. For this reason, the  ongoing surveys (e.g. \cite{Cameron17} for HTRU survey, \cite{Parent18} for PALFA survey, \cite{Morello20} for SUPERB survey)   also perform a  time-domain search with a Fast Folding Algorithm (FFA; \cite{Staelin69}), simultaneously to the frequency-domain periodicity search. 

\subsection{Targeted searches}
\label{subsec:bh3.2}
Targeted searches are more sensitive to pulsars compared to  wide-area surveys that cover the sky blindly. They allow deeper searches through longer observations (making the surveys more sensitive) as well as  multiple visits per source. This is precious  because in some cases a pulsar can be missed in a single observation due to scintillation, eclipses, or acceleration in a binary system. Such deep observations can characterise  specific environments in unique ways. Targeted surveys also probe different types of MSPs, so probing the evolutionary links between different classes.  
\subsubsection{Follow--up of high energy sources}
\label{subsec:bh3.2.1}
The radio and the high-energy ends of electromagnetic spectrum are highly complementary in pulsar searches, since  the highest sensitivity and resolution is attained in the radio domain but  the largest observable energy output (though weak in terms of photon counting statistics) is attained at higher energies (see Chapter~2).

\begin{figure}[t!]
\centering
\includegraphics[width=\textwidth]{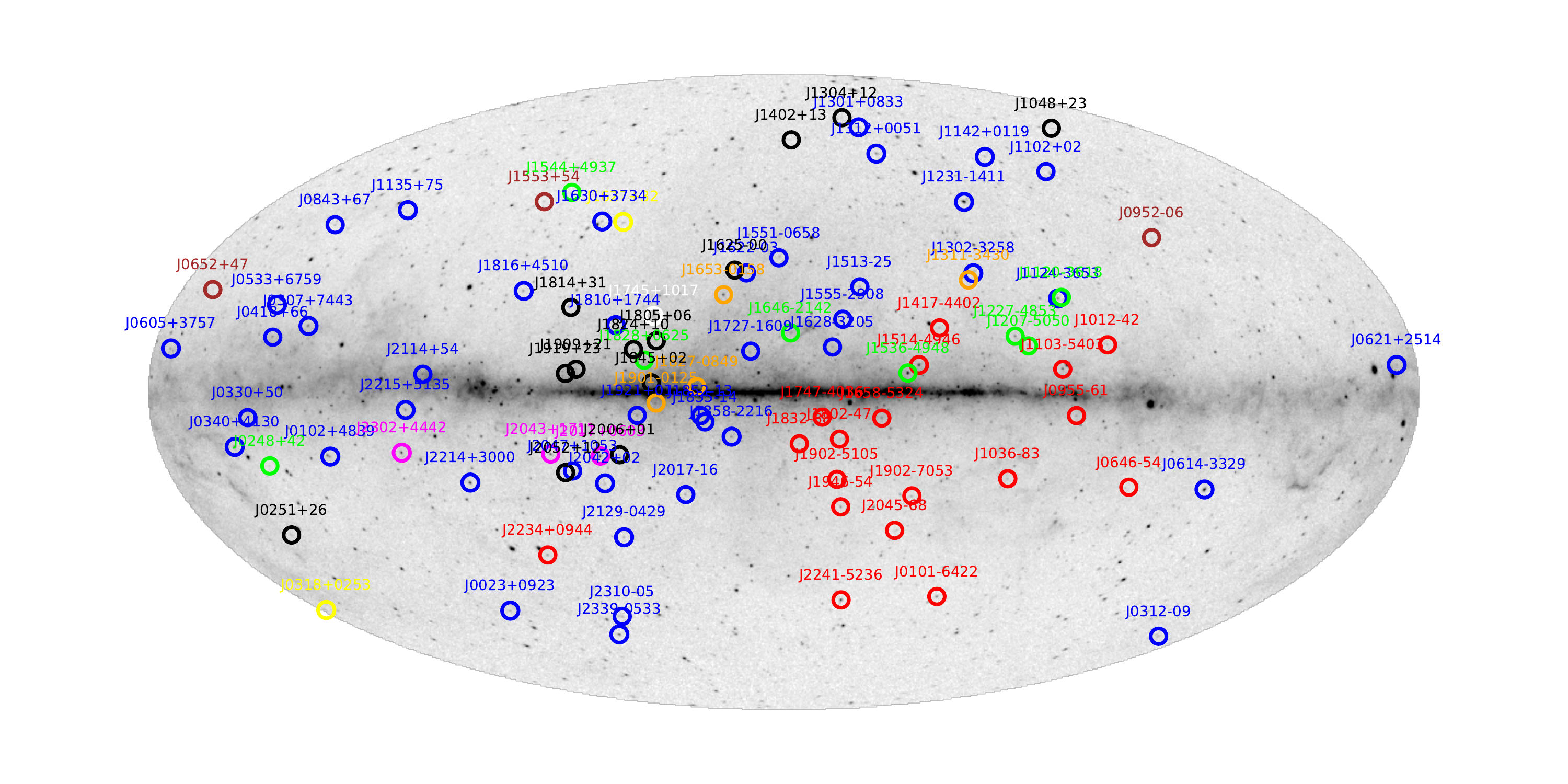}
\caption{Galactic distribution of the MSPs discovered in Fermi directed surveys. Discoveries by different telescopes are marked in different colours, 
GBT in blue, Parkes in red, GMRT in green, Nancay in magenta, white is Effelsberg,
orange is LAT, black is Arecibo, brown is LOFAR and yellow is FAST. Figure Courtesy: Paul Ray}
\label{fig_galactic MSP} 
\end{figure}

Targeted searches of high-energy  sources proved particularly efficient compared to blind surveys for pulsars,  especially so for the \emph{Fermi} directed searches. Since August 4, 2008, the Giga-election-volt $\gamma$-ray sky has been surveyed by the \emph{Fermi} Large Area Telescope ({\it LAT}, \cite{atwood09}), the primary instrument on-board the \emph{Fermi} Gamma-ray Space Telescope. An increasing number of unassociated $\gamma$-ray point sources appear at each \emph{Fermi} LAT catalog release. Targeted searches for radio pulsations at the position of such unassociated LAT point sources is coordinated by the \emph{Fermi} Pulsar Search  Consortium (PSC).  
Till now, 95 new MSPs are been discovered in this effort\footnote{https://confluence.slac.stanford.edu/display/GLAMCOG/Public+List+of+LAT-Detected+Gamma-Ray+Pulsars}, which amounts to about one third of the the total known Galactic MSP population. Fig. \ref{fig_cum_MSP} plots the cumulative number of known Galactic MSPs discovered over the years. It is evident  that Fermi directed surveys considerably enhanced  the number of MSPs. Fig.~\ref{fig_galactic MSP} shows the Galactic distribution of the MSPs discovered in Fermi directed surveys, and 
Table \ref{discovery} summarises the existing and ongoing efforts in Fermi-directed search for millisecond pulsation.

 Since MSPs are assumed of having been spun-up by accretion from a companion, many of them are naturally expected to be part of binary systems. LMXB/MSP transitional  systems are considered as direct observational evidences of this hypothesis, as it is described in detail in Chapter~6 of this book. On the other hand, the formation of isolated MSPs  has not been  understood yet. Their existence could be linked to a class of MSPs in very compact binary orbits (P$_b$ $<$ 1 day), known as spider MSPs, whose companion masses are either brown dwarfs (black widows, where 0.01 M$_{\odot}$$<$M$_c$$<$0.06 M$_{\odot}$) or main sequence stars (redbacks, where 0.1 M$_{\odot}$$<$M$_c$$<$0.8 M$_{\odot}$) \cite{Roberts13}. The eclipses of the radio signal observed from these systems are caused by plasma ejected by the radio pulsar wind and which surrounds the system. This process could evaporate completely the companion star of black widows, so producing isolated MSPs.  
The majority of spider MSP systems were discovered in the radio surveys of unassociated Fermi$-$ray sources.

\subsubsection{Searches in globular clusters}
\label{subsubsec:bh3.2.2}
\begin{figure}[t!]
\centering
\includegraphics[scale=.4,angle=-0]{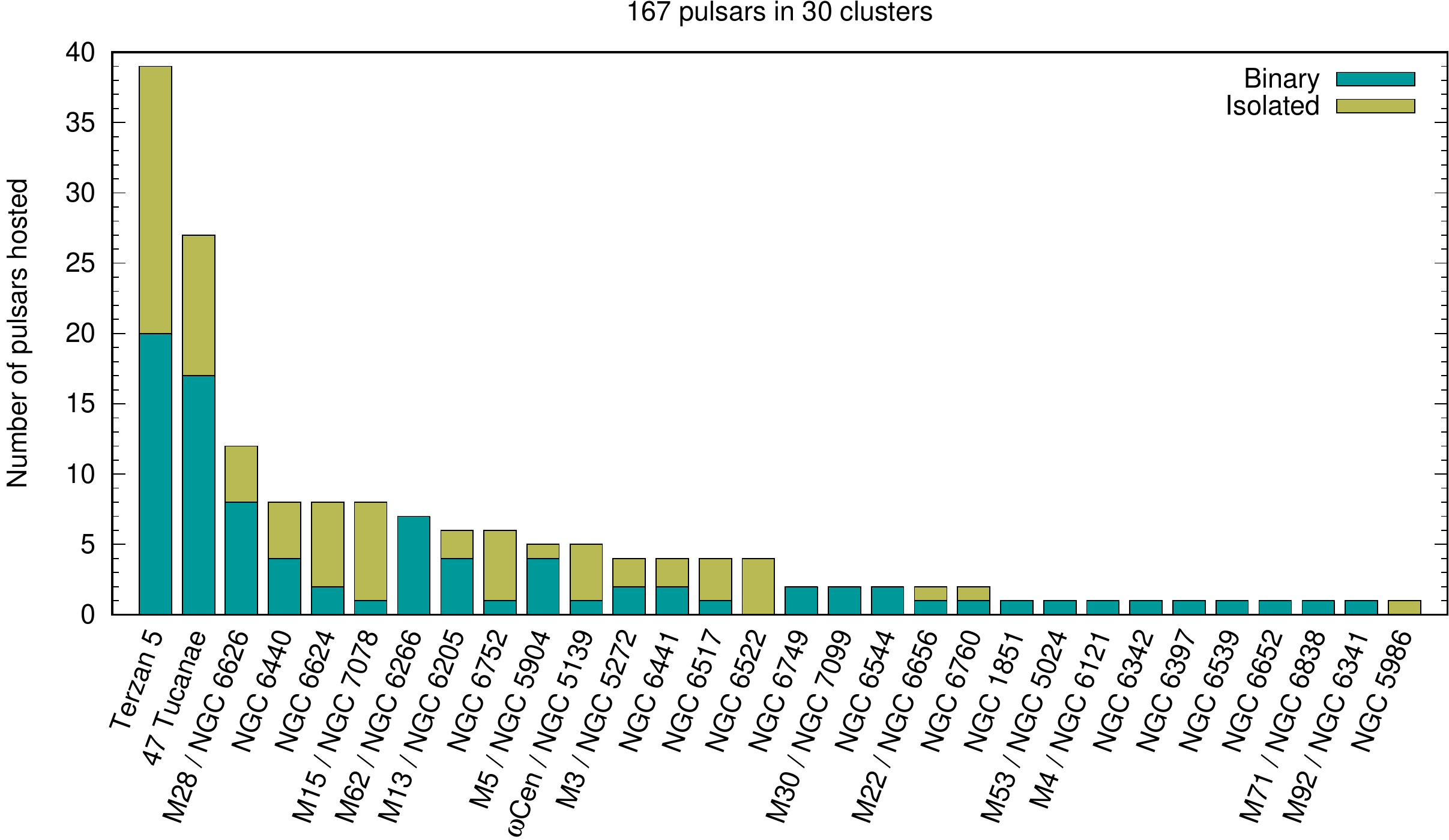}
\caption{Histogram of pulsars in globular clusters. Figure Courtesy: Alessandro Ridolfi}
\label{fig_globular_MSP} 
\end{figure}

Pulsars in globular clusters (GCs) are among the most interesting to study the pulsar population. From slow to millisecond pulsars, isolated to binary, GC pulsars have provided a variety of exciting science results. Over the last few decades, 157 pulsars have been discovered in 30 GCs, the vast majority of which are MSPs. Fig. \ref{fig_globular_MSP} plots the histogram of isolated and binary MSPs in individual globular clusters. Many of the GC MSPs are in exotic binary systems compared  to the population found in Galactic field. They include  the fastest spinning pulsar, highly eccentric binaries, and MSPs in peculiar evolutionary phases such as redbacks, black widows and a transitional MSP (e.g. \cite{Hessels06}, \cite{D'Amico01}, \cite{Ransom05}, \cite{Freire08}, \cite{DeCesar15}, \cite{Papitto13}). The high stellar densities in GCs lead to a high probability of close stellar interactions through which binaries form and subsequently evolve. GCs also hold the possibility of hosting ultra-fast spinning pulsars like sub-millisecond pulsars created through multiple episodes of recycling, which would  provide stringent constraints on the  neutron star equation-of-state \cite{Lattimer01}.    
Additionally, one can learn about  the cluster dynamics, gas content (e.g., \cite{Freire01}, \cite{D'Amico02}) and magnetic field, as well as about the MSP formation and evolution (thanks to the high number of sources; e.g., \cite{Rasio00}). The dense, highly interacting environment in clusters core could in principle host the rarest system, like a holy-grail MSP $-$ black hole binary or even an MSP $-$ MSP binary.

\label{subsec:bh3.2.1}

\subsubsection{Galactic Centre MSPs}
\label{subsubsec:bh3.2.3}
Scattering is the main hindrance in searches for MSPs in the Galactic centre region, since it causes a temporal broadening of pulses and a reduction of the pulse signal-to-noise ratio (S/N). The discovery of a 
MSP near the Galactic Centre region would be of immense importance, as it would be a very useful tool to measure the magnetised ISM in such extreme environment, and especially to probe the space-time surrounding the nearest super-massive black hole, Sgr A$^\ast$ \cite{liu12}. It is conjectured that General Relativity and other
perturbations experienced by Galactic Centre pulsars would be so large that even a low precision pulsar timing could suffice. Liu et al. \cite{liu12} illustrated that even with moderate timing precision, a binary MSP in a close orbit around Sgr A$\ast$ (P$_b$ $<$ 1 yr) would allow the observation of 
relativistic effects, like time dilation, Shapiro delay and relativistic peri-center precession.  It is expected that soon after the discovery of such an MSP, relativistic effects could be measured allowing an accurate measure  of the mass of Sgr A$\ast$ allowing unprecedented 
gravity tests. Pulsar timing has the potential of increasing the testing and diagnostic precision of gravity tests by at least three orders of magnitude In comparison with the best predicted precision of infrared astrometry and Doppler measurements.

\begin{figure}[t!]
\begin{center}
\includegraphics[scale=.5,angle=0]{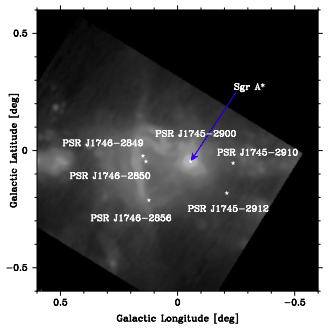}
\caption{The positions of known radio pulsars in the central 0.5$^{\circ}$ toward the GC overlaid on a 10.55 GHz
continuum map made with the Effelsberg telescope.  Future discovery of any MSP near to the Sgr A$^\ast$ will be very useful for gravity tests. Figure Courtesy B. Klein (Adapted from \cite{Eatough13}).}
\label{fig_gal_cen}
\end{center}
\end{figure}

To date, there are six active radio pulsars (all relatively slow {\it ordinary} pulsars) within 15$'$ ($\sim$36 pc) of Sgr A$^\ast$. Fig. \ref{fig_gal_cen} shows the distribution of these pulsars close to the Galactic Centre. This is a very small fraction of the  $\sim$ 1000 pulsars (inclunding MSPs) expected within the central pc around Sgr A$^\ast$ \cite{Wharton12, Chennamangalam14}.
Most of the  targeted searches of the Galactic Centre are performed at relatively high frequencies to minimise the loss of pulse power due to scattering, as the temporal broadening has a strong frequency dependence $\propto$ $\nu^{-4}$. However, pulsars become less luminous at such high frequencies because of their steep spectrum, which is described by a power-law spectrum with an index 
$\sim$ $-$1.7. The fact that a fraction of pulsars have relatively flatter spectral index, combined with the detection of a few pulsars, motivated high frequency searches in the Galactic Centre. The discovery of more pulsars towards the Galactic Centre would help determine the optimum frequency for targeted searches in the Galactic Centre.
For example, measurements of pulse broadening in PSR J1745$-$2900 found that the degree of scattering toward the Galactic Centre is less than previously predicted \cite{Spitler14}. This study also suggested that the scattering region could be distant from the Galactic Centre.

Periodicity searches that take into account the correction for the  for orbital motion (commonly known as acceleration and jerk searches; see Sect.~\ref{subsec:bh3.1}) are employed to search for millisecond pulsars in Galactic Centre region. Searches for transient sources are also relevant for the Galactic Centre region, because some pulsars may have strong single pulses or giant pulses, and are more easily detected by single-pulse detection algorithms.
\begin{figure}[t!]
\centering
\includegraphics[scale=.3,angle=-90]{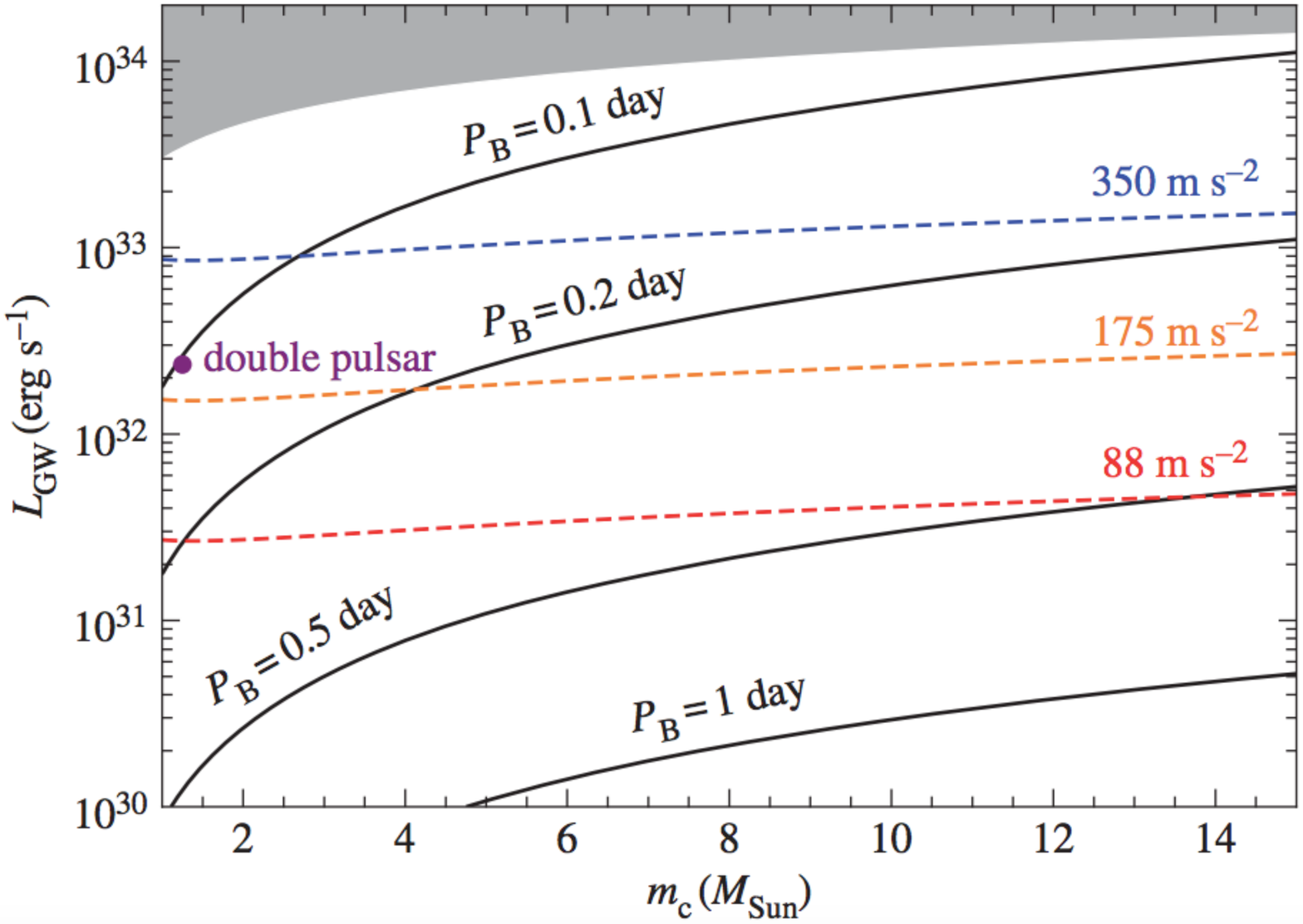}
\caption{The binary pulsar search space that the {\it SKA} will be sensitive to. A typical {\it SKA} observation will last about 540 s with an  acceleration search range up to 350 m $s^{-2}$. The higher GW luminosity for compact orbits (shorter P$_B$) with higher companion mass (m$_c$) gives a higher binary acceleration. Figure is taken from \cite{Stappers17}.}
\label{fig_SKA search} 
\end{figure}
\subsection{Wide field surveys}
\label{subsec:bh3.3}
The location of the majority of the pulsars cannot be predicted accurately. Thus, blind searches over large areas of the sky are the
best way to search for new pulsars.
There are a few factors that have to be considered in the design a blind survey, (i) the intrinsic steep spectra of MSPs (spectral index $\sim$ $-$1.7), (ii) the dispersion delays which scale as $\nu^{-2}$ and can be corrected, (iii) scattering, whose effect scales as  $\nu^{-4}$ and cannot be corrected, (iv) the sky temperature, and  (v) the field-of-view of the telescope. Low-frequency surveys are favoured as pulsars are intrinsically stronger, the field-of-view is larger, and scattering is relatively mild, especially for relatively higher galactic latitudes. 
Indeed, the most efficient all-sky pulsar searches are conducted within the radio frequency range between $\sim$ 100 MHz and $\sim$1 GHz, for Galactic latitudes $|b|>5^{\circ}$. 
Of the 396 MSPs listed in ATNF pulsar catalog, 35\% are at $|b|<5 ^{\circ}$, indicating the need for sensitive off-galactic plane surveys. 
The major, ongoing low-frequency pulsar surveys are listed in Table \ref{discovery}.  They are being conducted from 100 MHz to up till 1400 MHz with a sensitivity reaching up to 0.2 mJy. Considering the population model, these surveys will result in a significant number of interesting individual MSPs in the near future.

Pulsar searches with the future world's largest telescope, the Square Kilometre Array ({\it SKA})  featuring a wide field-of-view, high sensitivity, multi-beaming and sub-arraying capabilities, coupled with advanced pulsar search backends, aims to go ten times deeper than any ongoing wide-field survey resulting in discovering a large fraction of the Galactic pulsar population \cite{Keane15}. 
This will also increase the number  of MSPs suitable for high precision timing for the detection of gravitational waves (Sec. \ref{subsec:bh6.1}), double pulsars to test the theories of gravity (Sec. \ref{subsec:bh6.2}), and interesting individual system like triples (Sec. \ref{subsec:bh6.2}). The holy grail, a Pulsar-Black hole binary will be the long-sought-after system that the SKA is expected to find. The SKA1-Mid, which will be located in South Africa and operate from 350 MHz, is expected to find 700 MSPs (Fig. 1c of \cite{Stappers17}). The SKA1-Low, which  will be located in Australia and operate from 50 to 350 MHz, is expected to find a few hundreds nearby MSPs.  The SKA1-Low will have 500 tied-array beams over an observing bandwidth of 100 MHz, sampled at 100 $\mu$s, while the SKA1-Mid will produce 1500 tied-array beams over an observing bandwidth of 300 MHz sampled at 64 $\mu$s \cite{Levin17}. The SKA pulsar search backend can correct for 350 ms$^{-2}$ line-of-sight acceleration for a pulsar with 2 ms spin-period. For a typical SKA observations duration of 540 s, with this acceleration limit SKA can detect pulsar with a 15 M$_{\odot}$  black hole in a 5 hours orbit (seen in Fig. \ref{fig_SKA search} taken from \cite{Stappers17}). NS-NS or NS-WD binaries as compact as an hour could also be detected with the SKA.  

\section{Timing of MSPs}
\label{sec:bh4}
\begin{figure}[th]
\includegraphics[scale=0.8,angle=0]{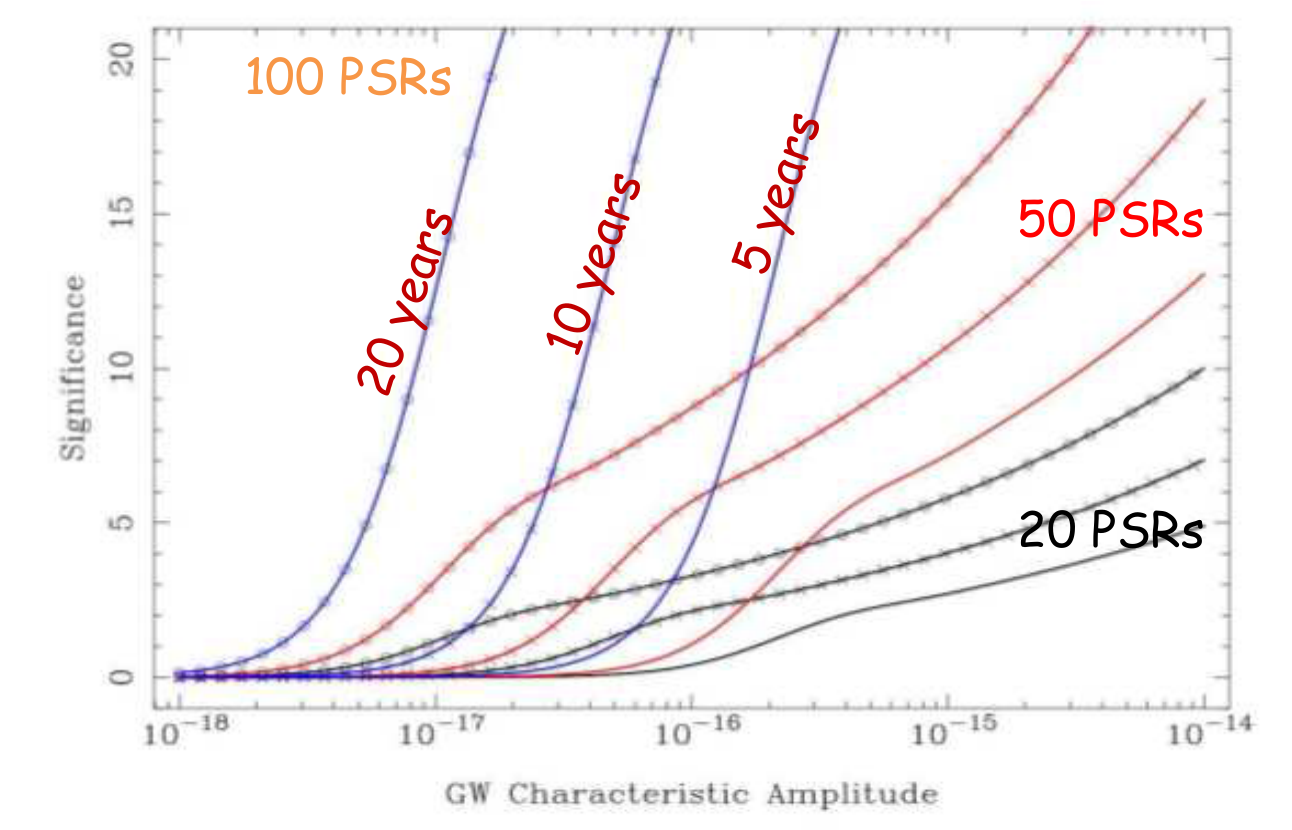}
\caption{Detection significance of the isotropic stochastic GW background using an ideal PTA, as a function of the signal strength, the number of MSPs in the PTA, and  data time span. In this simulation, a variable  number of MSPs is considered (20, 50 and 100), whereas the data span  5, 10 and 20 years, respectively. The RMS noise in the timing residuals is taken as 100 ns and a weekly cadence is assumed. 
For a GW amplitude of 10$^{-16}$, 5 years timing campaign with 100 pulsars can achieve higher significance than 20 years timing campaign with 20 pulsars. Figure taken from \cite{Manchester13}.}
\label{fig_PTA population} 
\end{figure}
Pulsar timing  probes interesting individual pulsar properties, like glitches \cite{Shemar96}, profile state changes \cite{Lyne10}, nulling \cite{Backer70}, intermittency  \cite{Kramer06a} and binary evolution. The timing of pulsars located in the Galactic plane also aids the investigation of properties of the interstellar medium through  the measure of scattering effects, as  well as via dispersion (DM) and rotation measure (RM) studies. Pulsar timing measures the deviation of the observed pulse arrival times from the values predicted with a timing model assumed beforehand. The measured deviations could be caused by more subtle un-modelled effects, like the low-frequency gravitational wave background or the curvature of the space-time in a compact massive binary. 
The high rotational stability \cite{Lorimer04}, compactness (second only to black holes), and their presence in binary systems, make MSPs ideal laboratories to test the physics of gravity \cite{Kramer06b,Cameron17,Stovall18}, to use them as detectors for long-wavelength gravitational waves \cite{Detweiler79,Foster90} and to constrain the equation of state of matter at supra-nuclear densities (\cite{Demorest10,Antoniadis13}; see the discussion in Chapters~9 and 10).
\begin{figure}[th]
\includegraphics[scale=0.3,angle=0]{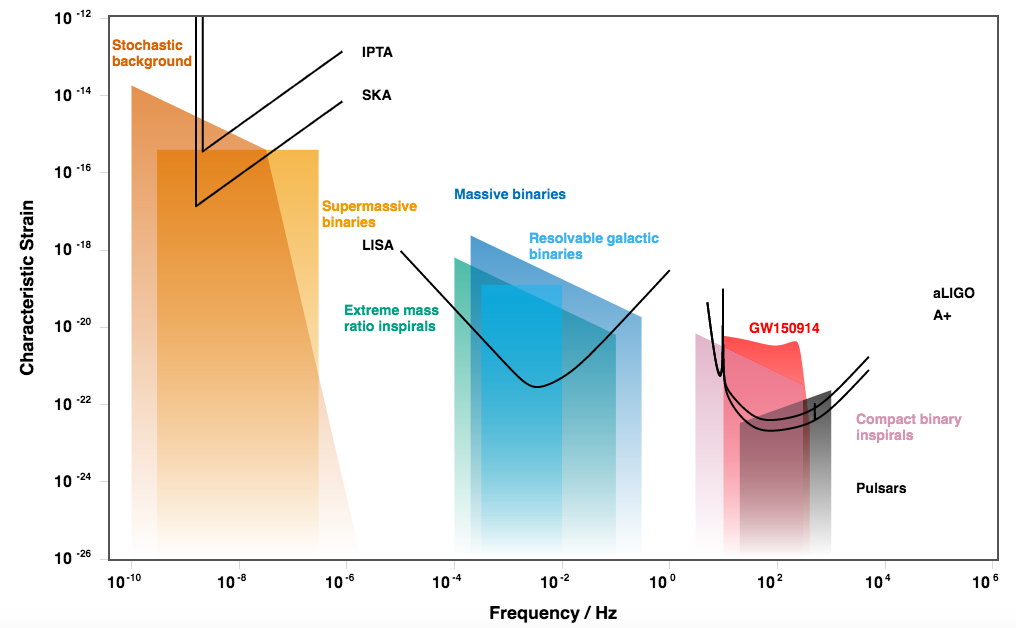}
\caption{Variation of the amplitudes of the gravitational wave radiation from variety of sources over the spectrum. Ground-based interferometers (e.g. LIGO and future instruments), space-based interferometers (e.g. LISA), and pulsar timing (e.g. IPTA and SKA) probe the spectrum fron nHz to kHz range from the cosmological population of SMBHBs to compact object inspirals. Produced with http://gwplotter.com/}
\label{fig_IPTA GWs} 
\end{figure}
\subsection{MSPs as sensitive gravitational wave detectors}
\label{subsec:bh6.1}
A Pulsar Timing Array (PTA) exploits the accurate pulse arrival time measurements of a large set of high-timing precision MSPs distributed across the sky, in order to detect the space-time vibrations caused by the stochastic Gravitational Waves (GWs) background \cite{Detweiler79,Foster90,Lee12}. As the timing span and precision increase, the timing residuals become dominated by red noise components, either from the GW background or from intrinsic pulsar timing noise. A significant fraction of the pulsar timing noise is produced by turbulence in the free electron density of the interstellar plasma. At present, one of the most crucial challenges for the PTAs is to disentangle the influence of the interstellar medium  from other sources of noise. Changes in the dispersion measure and/or the influence of scattering can adversely affect the arrival times. In order to separate the red noise components from the GW background, the non-white components of timing residuals due to the interstellar medium and/or the profile evolution with frequency need to be mitigated. Then, the detection of the stochastic GW background can be probed from the angular correlation between the residuals of the arrival times of pairs of pulsars distributed across the sky, which is called Hellings and Downs curve \cite{Hellings83}. Through these detections, a PTA aims at revealing the cosmic population of inspiralling supermassive black holes binaries (SMBHBs) in  merging galaxies. These systems are expected to emit GWs in the nanohertz frequency range, which is invisible to space-based  and ground-based GWs detectors like {\it LISA} and {\it LIGO} \cite{Spolaor18}. At present, there are three main PTA efforts: the EPTA (European Pulsar Timing Array\footnote{www.epta.eu.org}), NANOGrav (North American Nanohertz Observatory for GravitationalWaves\footnote{www.nanograv.org}) and the PPTA (Parkes Pulsar Timing Array\footnote{www.atnf.csiro.au/research/pulsar/ppta}). A PTA is sensitive to GWs with frequencies  ranging from 1/(timing-span) to 1/(cadence). Assuming a timing-span of 10 years and a weekly cadence, at the low frequencies (10$^{-9}$--10$^{-6}$ Hz) GWs are detectable through long-term timing observations of the most stable pulsars. All the three PTAs already provide stringent limits on the amplitude of the GW background around 10$^{-15}$, ruling out some of the theoretical models for the stochastic GW background \cite{Stappers17}. The timing measurements from the eight  telescopes used for the PTAs around the world are combined to form the International Pulsar Timing Array (IPTA) data set. This improves the cadence and the frequency range of the observations. 
In the recent second data release of the IPTA \cite{Perera19}, the combined high-precision timing data for 65 MSPs regularly observed by the three PTAs were considered. The significantly higher quality of these data promises to improve the limits on the isotropic stochastic low-frequency GW background obtained, so far. The GW landscape shown in Fig. \ref{fig_IPTA GWs} highlights the GW radiation expected from a variety of sources at frequencies ranging from nHz to kHz. The advent of {\it SKA} will increase the current limit 
of 10$^{-15}$ on the isotropic stochastic GW background by more than an order of magnitude. Besides improving the timing precision using sensitive instrument like the {\it SKA}, an increase in the number of MSPs that can be timed with good precision is also expected to contribute significantly to the PTA's sensitivity. According to Manchester et al. \cite{Manchester13}, for a GW amplitude of 10$^{-16}$, a 5 year-long timing campaign with an ideal PTA that uses 100 pulsars can achieve a higher significance than a 20 year-long campaign with 20 pulsars (see Fig. \ref{fig_PTA population}).

\subsection{MSPs as gravity probes}
\label{subsec:bh6.2}
MSPs in compact binaries with a white dwarf or a neutron star companion offer extreme environments to carry out stringent test of gravity. As the signal from the pulsar propagates through curved space-time across the binary, we can perform precision tests of gravity to search for tiny deviations from GR \cite{Kramer18}. 
In addition to the five Keplerian binary parameters like the orbital period P$_b$, the eccentricity of the orbit $e$, the projected semi-major axis of the pulsar orbit a$_{p}$sin$\it{i}$, the longitude of periastron $\omega$, and the epoch of periastron passage of the pulsar T$_0$, there are few “Post-Keplerian” (PK) parameters  which measure the relativistic effects in pulsar timing \cite{Lorimer04}. The PK parameters are the relativistic advance of the periastron described as a time derivative of the angle of periastron $\dot{\omega}$, the orbital decay due to gravitational wave emission described as time derivative of the orbital period $\dot{P_{b}}$, a combination of gravitation redshift and time dilation described by Einstein-delay parameter $\gamma$, a Shapiro-delay due to the curvature of space-time around the
companion described by a 'shape' s and a range 'r' parameter, and the relativistic deformation of the orbit described by $\delta _\theta$ and $\delta_r$ \cite{Lorimer04}. Different theories of gravity can be tested by comparing their predictions of the PK parameters for a binary system with those measured from timing studies. The Hulse-Taylor pulsar, B1913+16 \cite{Hulse75}  first provided an evidence for GW emission from the measure of a few PK parameters \cite{Taylor82}. Later, all the PK parameters could be measured for the exceptional double pulsar system, PSR J0737$-$3039A/B, so probing the deviations from GR by less than 0.05\% \cite{Kramer06b}. A recent study of two double neutron star systems (DNS) allowed the investigation of a hitherto unexplored relativistic parameter space while probing the formation and evolution of the DNS population. One of the DNS is in the most accelerated binary system PSR J1757$-$1854 \cite{Cameron17}, and the other is the most massive DNS system with the highest asymmetric mass ratio PSR J1913$+$1102  \cite{Ferdman20}. 
Also, PSR J0337$+$1715, a mildly relativistic hierarchical triple system with two white dwarfs, provided a unique laboratory to perform some of the most stringent tests of the strong equivalence principle (SEP), the universality of free fall \cite{Archibald18}. Finally, a yet to be found pulsar-black hole binary would be a holy-grail to trace the space–time around the most extreme gravitating system using the nature's best clocks. Finding such systems and performing high precision timing studies to reveal black hole properties with unprecedented sensitivities \cite{Liu14} is one of the main goal of the \emph{SKA}. 

\subsection{Eclipsing MSPs : Probing intra-binary material}
\label{subsec:bh6.3}
Black widow and redback spiders are eclipsing binary MSPs that can provide a variety of information on:\\ 
\begin{itemize}
\item the accretion history and the evolutionary link between accreting X-ray pulsars and radio MSP (see Chapter~6);\\
\item the high energy emission from the intra-binary shock produced by the  interaction between the pulsar and the stellar wind (see Chapter~2);\\
\item the eclipse mechanism and the properties of the eclipsing medium, such as magnetic field strength, temperature, clumping of material (described in this section and in Chapter~6).
\end{itemize}
Black widow and redback pulsars are fast spinning MSPs in compact binaries with a low-mass companion. These systems were named in this way as  it is assumed that the pulsar wind is energetic enough to ablate the companion star, or at least to eject the matter that the latter transfers to the neutron star and that is responsible for the observed eclipses of the radio signal. 
MSPs are naturally expected to be part of binary systems due to their formation in recycled scenario \cite{bh91}. Whereas the majority of the observed MSP systems are in binaries, a small fraction of these ($<$20\%) is isolated. It was argued that complete evaporation of the companion of a black widow MSP is one of the ways to form isolated MSPs \cite{alpar82}. Thus, these systems were proposed to be the evolutionary link connecting accreting X-ray pulsars and isolated millisecond pulsars. However, whether the companion in black widow systems system will be actually eventually evaporated is not clear yet. The mass loss rates determined by the recent studies for some of the systems indicates that it would be impossible to ablate the companion star completely within a Hubble time \cite{polzin19}.

\begin{figure}[th]
\includegraphics[scale=0.6,angle=0]{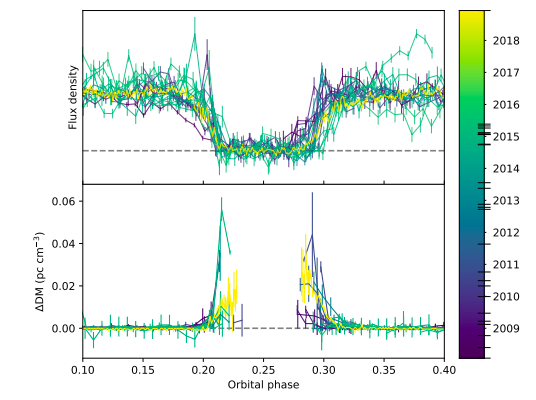}
\caption{Measured flux densities of PSR J2051$-$0827 for 345 MHz observations covering the eclipse region, with each normalised so the the out-of eclipse mean flux density is unity. The horizontal dashed line represents detection limit of the telescope. Bottom: Deviation from
mean out-of-eclipse dispersion measures for the same set of observations. 1$\sigma$ uncertainties from the simultaneous DM and scattering fits are shown with the error bars (Figure Courtesy : \cite{polzin19}).}
\label{fig_BW_MSP} 
\end{figure}

Both black widow and redback spider MSPs  have compact orbits with orbital periods ranging between 0.1$-$1 days;  black widows have very-low mass (0.01 $-$ 0.06 M$_{\odot}$) brown dwarf companions, whereas the redback systems have moderate mass (0.2 $-$ 0.8 M$_{\odot}$) main-sequence comapanion stars. As discussed in Section \ref{subsec:bh3.2.1}, the number of black widow and redback spider pulsars increased signficantly thanks to targeted pulsar searches  from $\gamma$-ray selected sources. Before the launch of Fermi, only two such eclipsing black widow systems  were known in the Galactic field,
 PSR B1957$+$20 \cite{Fruchter88} and PSR J2051$-$0827 \cite{Stappers96}. 
Presently there are $>$ 40 known black widow MSPs and 20 redback MSPs (with  parameters listed in the  catalogue maintained by A.~Patruno\footnote{https://apatruno.wordpress.com/about/millisecond-pulsar-catalogue/}).
Compared to other MSPs, the spider MSP systems are found to have on average higher values of spin-down energy-loss rate ($\dot{E}\sim 10^{34} erg~s^{-1}$
) making these systems good $\gamma$-ray pulsar candidates  emission \cite{Roberts13}. 

\begin{table*}
\caption{Parameters for eclipsing binary millisecond pulsar systems (adopted from Kudale et al. 2020).}
\begin{tabular}{l l l l l l}
\hline
Pulsar & Excess DM    & $\dot{E}/a^{2}$$^{\alpha}$ (10$^{35})$  & Eclipse  & n & Reference$^\zeta$ \\
Name   & (pc~cm$^{-3})$ & ($erg/s/R_{\odot}^{2})$ & duration$^{\beta}$ &   &\\
\hline
J1023$+$0038 (RB) & 0.15(700) &0.33 &40(685) & $-$0.41 & \cite{Archibald09}\\
 J1048+2339 (RB) & 0.008(327) & 0.03 & 57(327) & $-$ & \cite{Deneva16}\\
J1227$-$4853(RB) & 0.079(607) & 0.29 & 64(607) & $-$0.44    & \cite{kudale20}\\
J1227$-$4853$^{\gamma}$ (RB) & 0.035(607) & $-$ & 6(607) &   $-$ & \cite{kudale20}\\
J1544$+$4937 (BW) & 0.027(607) & 0.11 & 13(322) & $-$  & \cite{bh13}\\
J1723$-$2837 (RB) & $-$ & 0.04 & 26(1520) & $-$ & \cite{Crawford13}\\
B1744$-$24A (RB) &  0.6(1499.2) & $-$ & $\sim$50$^\delta$(820) & $-$ & \cite{Bilous19}\\
J1810$+$1744 (BW) & 0.015(325) & 0.18 & 13(149) & -0.41 & \cite{polzin18}\\
J1816$+$4510 (RB) &  0.01(149) & 0.08 & 24(121) &  -0.49$^\epsilon$ & \cite{polzin20}\\
 B1957$+$20 (BW) &  0.01(149) & 0.22 & 18(121) &  -0.18 & \cite{Bilous19}\\
J2051$-$0827 (BW)  &  0.13(705-4023) & 0.06 &  28(149) &  -0.41  & \cite{polzin19,polzin20}\\
J2215$+$5135 (RB) & $-$ & 0.28 & 66(149) & -0.21$^\epsilon$ & \cite{polzin19,Broderick16}\\
\hline
\end{tabular}
\\
Parameters presented in different columns of this table are,\\
Column 1: indicate redback (RB) or black widow (BW).\\ 
Column 2: the excess dispersion around eclipse boundary.\\ 
Column 3: $\dot{E}/a^2$, $\dot{E}$ is spin-down energy of the pulsar and $a$ is distance to the companion.\\  
Column 4: eclipse duration with corresponding frequency in parenthesis.\\ Column 5: index of power-law dependence (n) of full eclipse duration with frequency.\\
${\alpha}$: Using  https://apatruno.wordpress.com/about/millisecond-pulsar-catalogue/\\
${\beta}$: Eclipse duration (in \% of orbit) includes non-detection and
associated ingress, egress transition.\\
$\gamma$: Parameters for excess dispersion observed around inferior conjunction.\\
$\delta$: For majority of the observed eclipses. Eclipse duration are observed to be variable and sometimes completely enshrouding the pulsar \cite{Bilous19}.\\
$\epsilon$: Estimated power law index using all available frequency measurements from the recent literature.\\
 \label{tab:_bw_rb_literature_ref}
\end{table*}

The majority of the black widow and redback systems are eclipsed for a large fraction of the orbital period ($>$ 10\%, i.e., larger than the fraction of the orbit subtended by the companion’s Roche lobe). These eclipses are caused by the material of the very low mass companion blown by the pulsar wind. The energy flux of the isotropic pulsar wind of spider MSPs, at the distance of the companion, is at-least four order of magnitudes larger  than other MSPs ($\dot{E}/a^2$, where $a$ is the distance to the companion,
for the black widow and redback systems is $\sim$ 10$^{34}$ $erg/s/R_{\odot}^{2}$, whereas $\dot{E}/a^{2}$ for other MSPs are $10^{29} -10^{30}erg/s/R_{\odot}^{2}$). Table \ref{tab:_bw_rb_literature_ref}
lists the observed properties of black widow and redback MSPs.
A higher value of $\dot{E}/a^{2}$ indicates that an active interaction between the pulsar and its companion is likely. 
During the eclipse, the radio flux density  regularly drops below the detection threshold throughout the companion superior conjunction, with an excess of DM  near the eclipse boundaries. Fig. \ref{fig_BW_MSP} shows the flux density observed for the black widow MSP system PSR J2051$-$0827
\cite{polzin19}. It is shown that at 345 MHz the MSP is never detected between orbital phases, 0.23$-$0.27, with a corresponding increase of the DM value near the eclipse boundary.

Thompson et al. \cite{thompson94} described possible mechanisms for the observed eclipses. Most of the observational studies of the black widow and redback spider systems pointed to cyclo-synchrotron absorption as the most likely eclipse mechanism (e.g. PSR B1957$+$20 \cite{faucher06}, PSR J2051$-$0827 \cite{Stappers96,polzin19,polzin20}, PSR J1810$+$1744 \cite{polzin18}, J1227$-$4853 \cite{roy15,kudale20}). While the earlier studies used only timing observations (e.g.\cite{faucher06,Stappers96}), some of more recent studies (e.g.\cite{roy15,polzin19,kudale20}) used simultaneous timing and imaging observations to probe the eclipse mechanism. 
In addition, polarisation study at the eclipse boundary has been performed \cite{Crowter20,Li19,polzin19}, resulting in measurement of the magnetic field in the eclipse region.
In addition to the main eclipse, random short duration eclipses have also been reported from a few spider MSPs
(e.g., in the case of PSR J1544$+$4937, \cite{bh13}). A recent study by Kudale et al. \cite{kudale20} has reported a flux decrease near the pulsar inferior conjunction of PSR J1227$-$4853. This can be interpreted as caused by mass loss through the L2 Lagrangian point for a system that has a rapid orbital evolution.

\end{document}